\documentclass[aoas,MSNbibl,nameyear,dvips]{arximspdf}

\doi{10.1214/13-AOAS614C} 
\referstodoi{10.1214/12-AOAS614}
\volume{7}
\issue{4}
\pubyear{2013}
\firstpage{1876}
\lastpage{1880}

\makeatletter
\renewcommand{\citep}[1]{(\citeauthor{#1} \citeyear{#1})}
\makeatother

\begin{document}
\begin{frontmatter}

\title{Discussion of ``Estimating the historical and future
probabilities of large terrorist events'' by~Aaron~Clauset~and~Ryan~Woodard}
\runtitle{Discussion}

\begin{aug}
\author{\fnms{Gentry} \snm{White}\corref{}\ead[label=e2]{gentry.white@qut.edu.au}}
\runauthor{G. White}
\affiliation{Queensland University of Technology}
\address{Science and Engineering Faculty\\
Mathematical Sciences School\\
Queensland University of Technology\\
Brisbane, Queensland 4001\\
Australia\\
\printead{e2}}
\pdftitle{Discussion of ``Estimating the historical and future
probabilities of large terrorist events'' by Aaron Clauset and Ryan Woodard}
\end{aug}

\received{\smonth{7} \syear{2013}}
\revised{\smonth{8} \syear{2013}}



\end{frontmatter}
%
\section{Introduction}

The terrorist attacks in the United States on September 11, 2001
appeared to be a harbinger of increased terrorism and violence in the
21{st} century, bringing terrorism and political
violence to the forefront of public discussion.
%
Questions about these events abound, and ``Estimating the Historical
and Future Probabilities of Large Scale Terrorist
Event'' [\citet{Clauset:Woodard:2013}] asks specifically, ``how rare are
large scale terrorist events?'' and, in general,
encourages discussion on the role of quantitative methods in terrorism
research and policy and decision-making.


%
Answering the primary question raises two challenges. The first is
identifying terrorist events.
The second is finding a simple yet robust model for rare events that
has good explanatory and predictive capabilities.
%
The challenges of identifying terrorist events is acknowledged and
addressed by reviewing and using data from two well-known and reputable
sources: the Memorial Institute for the Prevention of Terrorism-RAND
database (MIPT-RAND) [\citeauthor{MIPT}] and the Global Terrorism Database
(GTD) [\citet{GTD:2012,LaFree:Dugan:2007}].
\citet{Clauset:Woodard:2013} provide a detailed discussion of the
limitations of the data and the models used, in the context of the
larger issues surrounding terrorism and policy.

The models proposed fit tail probabilities for power-law and
alternative models based on data from both the MIPT-RAND database and
the GTD. These models are thoroughly explained and well executed, as
are the results.
The predictive capabilities and forecasts, along with consideration of
the influence of exogenous factors such as attack type, target and
economic development are considered, presented and discussed clearly,
affirming the robustness of the methods. The authors estimate that, in
the 40-year period since 1968, there is an 11--35\% chance of a terror
event at least the size of September 11, 2001. 

\section{Comments}
Terrorism and political violence
are complex phenomenon of human behavior [\citet
{Horgan:Boyle:2008,Taylor:Horgan:2006}], and
rely on the fear and uncertainty surrounding rare events to create a
disproportionate effect that is difficult to directly measure [\citeauthor{Crenshaw:1986}
(\citeyear{Crenshaw:1981,Crenshaw:1986}), \citet{Waugh:1983,,Thorton:1964}]. In this
context, making predictions about human behavior is a tricky business,
and interpreting an 11--35\% probability for an extreme event
illustrates part of this problem.
An 11--35\% probability sounds ominous, but, over a 40-year period,
that translates to a seemingly innocuous daily probability around 1 in 100,000.
The temptation is to interpret an 11--35\% chance as near certainty, and
a 1 in 1,000,000 chance as near impossibility.
Neither of these time scales for interpretation are useful, and belie a
further problem with considering large scale historical trends when
making predictions about rare events using only previous history.
%


For example, in 224 years there have been 44 US Presidents; 4 (9\%)
were assassinated. The first was in 1865; in the period between 1865
and 1901, 3 of the 10 US Presidents were killed in office (30\%). Since
1901, only 1 (5\%) US President was assassinated. Making a forecast in
1864, and relying solely on historical data, the expected number of US
Presidents killed in office in the ensuing 40 years would be 0. In
1902, looking back at the previous 40 years, there would be an expected
5 US Presidents killed in office during the 20{th}
century. This example is hardly definitive, but it illustrates the
point that rare events involving humans are difficult to predict.

While \citet{Clauset:Woodard:2013} (rightfully) do not address this,
their paper does address the caveats of its results in great detail,
which provides the basis for raising the question, ``what is the role
of quantitative methods in terrorism research and in assisting policy
and decision makers?''

In \citet{Lum:2006}, a systematic review of the literature reveals a
significant increase
in research on terrorism and counterterrorism efforts since 2001,
though only a small minority apply quantitative methods. 
Despite this, there are some notable examples, both included in \citet
{Lum:2006} and after.
\citet{Enders:Sandler:1993} use vector autoregression (VAR) and an
interrupted time series approach to model the effects of
counterterrorism policies on transnational terrorism from 1968 through
1988. \citet{Dugan:etal:2005} and \citet{Dugan:2011} use Cox proportional
hazard models, and their variants, to analyze the effects of
interventions on hijackings and IRA terrorist activity in Northern
Ireland [\citet{Lafree:etal:2009}].
%
\citet{Arce:Sandler:2005} propose a game theoretic framework for
modeling the interactions between terrorists and counterterrorism
efforts, and \citet{Sapperstein:2008} and \citet{Minami:2009} suggest
modeling the interaction between terrorists and counterterrorism
efforts using a dynamic linear modeling approach.
%
Recent research shows that patterns of terrorist activity are well
modeled using a cluster process interpretation of self-exciting process
models [\citeauthor{Hawkes:1971a} (\citeyear{hawkes71b,Hawkes:1971a}), \citet{hawkesoakes:1974}].
Self-exciting models have been applied to airline hijackings [\citeauthor{Holden:1986}
(\citeyear{Holden:1986,Holden:1987})], insurgent activity in Iraq [\citet
{Mohler:2010,Mohler:etal:2011,Lewis:etal:2011,
Lewis:Mohler:2012}] and terrorism data from Southeast Asia and Colombia
[\citet{Porter:White:2012,White:etal:2012b,White:Porter:2013}].

One important aspect of modeling terrorism that is not explicitly
stated, but is implicit in \citet{Clauset:Woodard:2013}, is the notion
of different processes for different levels and types of terrorist
activity. This idea, illustrated by the fitting of tail probabilities
for rare events, can help explain the relative scarcity and sporadic
nature of terrorism [\citet{Porter:White:2012,Vasanthan:etal:2013}].
This extends to a complex, unobserved latent process as a model for the
occurrence, and resulting characteristics, of terrorist events. The
capability to model and describe complex unobserved processes is well
established and is an ongoing area of significant research in the
mathematical and statistical sciences.
%
The advent of newly available data sources, like the GTD and the MIPT
data sets, and increased awareness outside of the field of terrorism
studies creates an opportunity for mathematicians and statisticians to
work more closely and in conjunction with experts from academia, in
policy and decision-making roles to create new models and methods to
expand our understanding of terrorism and terrorist activity.

For the quantitative researcher, the utility of these models is
obvious. As exploratory tools they can reveal heretofore unobserved
patterns in activity. As confirmatory tools they can be used to test
specific ideas and theories about these patterns. The challenge for the
quantitative researcher is to understand their place in the field of
terrorism studies as a whole, assisting in the building of sound
knowledge, and aiding policy and decision makers.

\section{Conclusion}
Terrorism studies itself faces an important epistemological quandary,
and there is an ongoing debate over whether terrorism---however it is
understood---should be analyzed within its individual context or
whether it can be assessed on a more universal level, across space and time
[\citet{Silke:2001,Weinberg:etal:2004,Duvesteyn:2004,Neumann:2009}].
As a result, the role of quantitative (particularly statistical)
methods in terrorism studies is often lost in this debate.
The argument of terrorism scholars is that individual terrorists and
acts of terrorism are too unique to benefit from statistical analysis.
The statistical perspective is that the purpose of the statistical
analysis of data is to make inferences about the underlying process or
\emph{context} that produce the data, not specific observations.
Or, in the words of Sherlock Holmes:

\begin{quote}
``\ldots while the individual man is an insoluble puzzle, in the aggregate
he becomes a mathematical certainty. You can, for example, never
foretell what any one man will do, but you can say with precision what
an average number will be up to. Individuals vary, but percentages
remain constant.'' The Sign of Four---Sir Arthur Conan Doyle [\citet{Doyle:2000}]
\end{quote}

While Holmes is a fictional character, his statement neatly sums up the
miscommunication that often occurs between statisticians--nonstatisticians.
Terrorism studies scholars and policy and decision
makers want (and rightly so) predictions at a very fine level, down to
the individual's behaviors. Statisticians should agree that this is
often beyond the reasonable expectation of their methods. But
statistical and
quantitative methods can contribute understanding on combating
terrorism by identifying and measuring specific differences \emph{between}
contexts (i.e., countries, regions or groups). These can be analyzed to
identify contextual differences and explore \emph{why} they exist,
providing a deeper understanding of terrorism and political violence.

Statistical methods do not intend to provide definitive answers; their
results, couched in uncertainty, should inform, not make decisions.
In order to advance the understanding of terrorism, the benefits and
limitations of quantitative methods need to be clearly understood, and
it is the role and duty of the expert to clearly and effectively
communicate the benefits and limitations of quantitative methods to
qualitative researchers and policy and decision makers.

\section*{Acknowledgment} The author would like to extend his
appreciation to Ms. Kate Irwin, ML for providing excellent feedback and
suggestions to improve this manuscript.

%



\printaddresses

\end{document}